 \documentclass[tablecaption=bottom,wcp]{jmlr} 

\usepackage{booktabs}
\usepackage{xcolor}
\usepackage[load-configurations=version-1]{siunitx} 
\usepackage{enumitem}

\newenvironment{proof_sketch}{\textbf{Proof Sketch\hspace*{4pt}}}{\hfill$\blacksquare$}

\jmlrproceedings{AABI 2022}{4th Symposium on Advances in Approximate Bayesian Inference, 2022}

\title[Metropolis Augmented Hamiltonian Monte Carlo]{Metropolis Augmented Hamiltonian Monte Carlo}

\author{\Name{Guangyao Zhou} \Email{stannis@vicarious.com}\\
  \addr Vicarious AI}

\begin{document}

\maketitle

\begin{abstract}
Hamiltonian Monte Carlo (HMC) is a powerful Markov Chain Monte Carlo (MCMC) method for sampling from complex high-dimensional continuous distributions. However, in many situations it is necessary or desirable to combine HMC with other Metropolis-Hastings (MH) samplers. The common HMC-within-Gibbs strategy implies a trade-off between long HMC trajectories and more frequent other MH updates. Addressing this trade-off has been the focus of several recent works. In this paper we propose Metropolis Augmented Hamiltonian Monte Carlo (MAHMC), an HMC variant that allows MH updates within HMC and eliminates this trade-off. Experiments on two representative examples demonstrate MAHMC's efficiency and ease of use when compared with within-Gibbs alternatives.
\end{abstract}

\section{Introduction}

Hamiltonian Monte Carlo (HMC) is a popular Markov Chain Monte Carlo (MCMC) method. It samples from complex high-dimensional distributions $\pi(q)\propto \exp(-U(q))$ on continuous variables $q\in \mathcal{R}^{n}$, where $U(q):\mathcal{R}^{n}\to \mathcal{R}$ is commonly referred to as the potential energy. HMC has enjoyed remarkable empirical success, due to the use of powerful symplectic integrators~\citep{leimkuhler2004simulating} (e.g. the leapfrog integrator) to maintain high acceptance probabilities for long-range gradients-guided proposals~\citep{neal2012mcmc, betancourt2017conceptual}. Common HMC implementations involve simulating the Hamiltonian dynamics for multiple leapfrog steps, followed by a Metropolis Hastings (MH) correction. \cite{chen2020fast} recently rigorously establishes the importance of using multiple leapfrog steps/long trajectories for HMC's efficiency, especially when compared with the Metropolis Adjusted Langevin Algorithm (MALA)~\citep{dwivedi2019log}, a special case of HMC using only one leapfrog step and a widely used algorithm in Bayesian statistics and machine learning.

However, we often face distributions of the form $\pi(q^{H}, q^{O}) \propto \exp(-U(q^{H}, q^{O}))$, where we can only use HMC for the continuous variables $q^{H}\in \mathcal{R}^{n}$, and it is necessary or desirable to use some other MH samplers for the variables $q^{O}$ (Sec.\ 4.3 of~\cite{neal2012mcmc}). Some common situations include (1) when $q^{O}$ are discrete, (2) when we have specialized MH samplers for $q^{O}$ that are efficient/easy to use, or (3) when $q^{O}$ are continuous but $\nabla_{q^{O}} U(q^{H}, q^{O})$ is expensive or impossible to compute. In such cases, we typically adopt an HMC-within-Gibbs strategy~\citep{neal2012bayesian, dang2019hamiltonian, kelly2021directly}, where we alternate between HMC updates and other MH updates. However, this implies a trade-off between long HMC trajectories and more frequent other MH updates. Longer HMC trajectories can help suppress random walk behavior, but might hurt overall sampling since other MH updates can only be done infrequently (between HMC updates). Shorter HMC trajectories allow more frequent other MH updates, but lead to increased random walk behavior.

Several recent works have focused on the above trade-off. \cite{neal2020non} combines MALA with partial momentum refreshment~\citep{horowitz1991generalized} and non-reversible Metropolis accept/reject decisions to allow more frequent other MH updates while suppressing random walk behavior. \cite{zhou2020mixed, zhou2021erratum} proposes mixed HMC (M-HMC) for distributions with mixed discrete and continuous variables to allow making discrete MH updates as part of an HMC trajectory. In this paper, we propose Metropolis Augmented Hamiltonian Monte Carlo (MAHMC) to completely eliminate this trade-off. MAHMC interprets HMC as a deterministic MH proposal, and augments HMC by introducing MH updates as part of the HMC trajectory. MAHMC is generally applicable, can be trivially implemented on top of HMC with no additional overhead, and includes M-HMC with Laplace momentum as a special case. We demonstrate MAHMC's advantage over within-Gibbs alternatives~\citep{neal2020non} and ease of use on two representative examples in Sec.~\ref{sec:exps}.

\begin{algorithm2e}[t!]
\caption{Basic components for sampling from $\pi(q)\propto \exp(-U(q)), q\in \mathcal{R}^{n}$}
\label{algo:components}
\SetKwProg{Def}{def}{}{end}
\SetKwFunction{leapfrog}{leapfrog}
\Def{\leapfrog$(q, p, \epsilon | U)$}{
	$p \leftarrow p - \frac{\epsilon}{2}\nabla U(q)$; $q \leftarrow q + \epsilon p$; $p \leftarrow p - \frac{\epsilon}{2}\nabla U(q)$;\\
	\KwRet $q, p$;
}
\SetKwFunction{mh}{MH\_correction}
\Def{\mh$(q_0, p_0, q, p, v | U, K)$}{
$E_0 \leftarrow U(q_0) + K(p_0)$; $E \leftarrow U(q) + K(q)$;\\
\uIf{$|v| \le   \exp(-E + E_0)$}{$v \leftarrow v \exp(-E_0 + E)$}
\lElse{$q, p \leftarrow q_0, p_0$}
\KwRet $q, p, v$;
}
\SetKwFunction{multipleleapfrog}{multiple\_leapfrogs}
\Def{\multipleleapfrog$(q, p, \epsilon, L | U)$}{
\lFor{$1$ \KwTo $n$}{
	$q, p \leftarrow$ \leapfrog$(q, p, \epsilon | U)$
}
$p \leftarrow -p$;\\
\KwRet $q, p$;
}
\end{algorithm2e}
\vspace*{-4pt}

\section{Background}

\subsection{HMC}\label{sec:hmc}

HMC introduces auxiliary momentum $p\in \mathcal{R}^{n}$ associated with kinetic energy $K(p)=\sum_{i=1}^{n} p_i^{2}/2$, and simulates the Hamiltonian dynamics for $L$ steps of size $\epsilon$  using the leapfrog integrator (\textit{leapfrog} in Algo.~\ref{algo:components}), before making the final MH correction (\textit{HMC} in Algo.~\ref{algo:algorithms}). We observe  that we can in fact interpret the multiple leapfrog steps (\textit{multiple\_leapfrogs} in Algo.~\ref{algo:components}) as a deterministic MH proposal, and derive the final MH correction as the usual MH acceptance probability for the deterministic proposal (Sec. 5.2 of~\cite{betancourt2017conceptual}).

\subsection{MALA and MALA variants}\label{sec:mala}

Using $L=1$ leapfrog step in HMC results in the widely used special case MALA. MALA uses gradients information, and allows more frequent other MH updates, but suffers from random walk behavior due to the use of short trajectories with frequent momentum refreshments.  Partial momentum refreshment~\citep{horowitz1991generalized} (MALA-P in Algo.~\ref{algo:algorithms}) was proposed as a possible remedy. However, as~\cite{neal2020non} explains, a rejection would lead MALA-P to almost double back on itself, making it less efficient than HMC with long trajectories.

\cite{neal2020non} proposes a non-reversible scheme for Metropolis accept/reject decisions (MALA-PN in Algo.~\ref{algo:algorithms}) to maintain the ability to make frequent MH updates while further suppressing the random walk behavior that comes from MALA-P doubling back on itself due to rejections. MALA-PN produces long rejection-free runs by encouraging rejections to cluster together, and demonstrates improved performance on multiple problems. 

\begin{algorithm2e}[t!]
\caption{Algorithms for sampling from $\pi(q)\propto \exp(-U(q)), q\in \mathcal{R}^{n}$}
\SetNoFillComment
\label{algo:algorithms}
\SetKwProg{Def}{def}{}{end}
\SetKwFunction{mh}{MH\_correction}
\SetKwFunction{multipleleapfrog}{multiple\_leapfrogs}
\SetKwFunction{hmc}{HMC}
\Def{\hmc$(q_0,\epsilon, L | U, K)$\tcp*[f]{Sec.~\ref{sec:hmc}}}{
$p_0 \sim N(0, I_{n})$; $q, p \leftarrow q_0, p_0$;\\
$q, p \leftarrow$ \multipleleapfrog$(q, p, \epsilon, L | U)$; \tcp*[f]{Defined in Algo.~\ref{algo:components}}\\
$q, p, v \leftarrow$ \mh$(q_0, p_0, q, p, \textit{Uniform}(0, 1) | U, K)$;\tcp*[f]{Defined in Algo.~\ref{algo:components}}\\
\KwRet $q$;
}
\SetKwFunction{partial}{MALA-P}
\Def{\partial$(q_0, p_0, \epsilon, \alpha | U, K)$\tcp*[f]{Sec.~\ref{sec:mala}}}{
$n \sim N(0, _{n})$; $p_0 \leftarrow \alpha p_0 + \sqrt{1 - \alpha^2} n $; $q, p \leftarrow q_0, p_0$;\\
$q, p \leftarrow$ \multipleleapfrog$(q, p, \epsilon, 1 | U)$;\tcp*[f]{Defined in Algo.~\ref{algo:components}}\\
$q, p, v \leftarrow$ \mh$(q_0, p_0, q, p, \textit{Uniform}(0, 1) | U, K)$;\tcp*[f]{Defined in Algo.~\ref{algo:components}}\\
\KwRet $q, -p$;
}
\SetKwFunction{partialnonreversible}{MALA-PN}
\Def{\partialnonreversible$(q_0, p_0, v, \epsilon, \alpha, \delta | U, K)$\tcp*[f]{Sec.~\ref{sec:mala}}}{
$n \sim N(0, I_{n})$; $p_0 \leftarrow \alpha p_0 + \sqrt{1 - \alpha^2} n $; $q, p \leftarrow q_0, p_0$;\\
$q, p \leftarrow$ \multipleleapfrog$(q, p, \epsilon, 1 | U)$;\tcp*[f]{Defined in Algo.~\ref{algo:components}}\\
$q, p, v \leftarrow$ \mh$(q_0, p_0, q, p, v | U, K)$;\tcp*[f]{Defined in Algo.~\ref{algo:components}}\\
$v \leftarrow (v + 1 + \delta)\bmod 2 - 1$; \\
\KwRet $q, -p, v$;
}
\end{algorithm2e}
\vspace*{-2pt}

\subsection{Within-Gibbs sampling from $\pi(q^{H}, q^{O})\propto \exp(-U(q^{H}, q^{O})), q^{H} \in \mathcal{R}^{n}$}\label{sec:within-gibbs}

For distributions $\pi(q^{H}, q^{O}) \propto \exp(-U(q^{H}, q^{O}))$, $q^{H} \in \mathcal{R}^{n}$ where we want to use MH updates for $q^{O}$, we refer to the common strategy of alternating between updates of $q^{H} \in \mathcal{R}^{n}$ (using a sampler for continuous distributions) and MH updates of the other variables $q^{O}$ as within-Gibbs sampling. In our experiments, we use four types of within-Gibbs samplers as baselines: (1) MALA within Gibbs (MALAwG), (2) HMC within Gibbs (HwG), (3) MALA-P within Gibbs (MALA-PwG), and (4) MALA-PN within Gibbs (MALA-PNwG).

\subsection{M-HMC for distributions with mixed discrete and continuous variables}

\cite{zhou2020mixed, zhou2021erratum} proposes M-HMC, an HMC variant that evolves the discrete and continuous variables in tandem for distributions with mixed support. M-HMC naturally supports frequent MH updates within long HMC trajectories, and demonstrates improved performance over strong baselines. In Sec.~\ref{sec:mhmc-connection}, we show that the practically useful M-HMC implementation (with Laplace momentum) can be seen as a special case of MAHMC.

\begin{algorithm2e}[t!]
\caption{MAHMC. \textcolor{blue}{Blue} highlights changes on top of naive MH within HMC.}
\label{algo:mahmc}
\SetKwProg{Def}{def}{}{end}
\SetKwFunction{mahmc}{MAHMC}
\SetKwFunction{leapfrog}{leapfrog}
\Def{\mahmc$(q^{H}_0, p^{H}_{0}, q^{O}_0,  \epsilon, L | U, K, \mathbb{Q}_{i}, i=1, \ldots, N^{O}, \mathbb{P}^{\mathcal{D}})$\tcp*[f]{Sec.~\ref{sec:mahmc-algo}}}{
	$q^{H}, p^{H}, q^{O} \leftarrow q^{H}_0, p^{H}_0, q^{O}_0$; \textcolor{blue}{$D \sim \mathbb{P}^{\mathcal{D}}(\cdot)$; $\Delta E \leftarrow 0$;}\\
	\For{$j \leftarrow 1$ \KwTo $L$}{
		\uIf{$D_j = 0$}{
			$q^{H}, p^{H} \leftarrow \leapfrog(q^{H}, p^{H}, \epsilon | U(\cdot, q^{O}))$;\tcp*[f]{Defined in Algo.~\ref{algo:components}}\\
		}
		\uElse{
			$\tilde{q}^{O} \sim \mathbb{Q}_{D_{j}}(\cdot| q^{H}, q^{O})$;\\
			\uIf{$\textit{Uniform}(0, 1) \le  \frac{\exp(-U(q^{H}, \tilde{q}^{O}))\mathbb{Q}_{D_{j}}(q^{O} | q^{H}, \tilde{q}^{O})}{\exp(-U(q^{H}, q^{O}))\mathbb{Q}_{D_{j}}(\tilde{q}^{O} | q^{H}, q^{O})}$}{
				$q^{O} \leftarrow \tilde{q}^{O}$; \textcolor{blue}{$\Delta E \leftarrow \Delta E + U(q^{H}, \tilde{q}^{O}) - U(q^{H}, q^{O})$}; \\
			}
		}
	}
	$E \leftarrow U(q^{H}, q^{O}) + K(p^{H})$; $E_{0} \leftarrow U(q^{H}_0, q^{O}_0) + K(p^{H}_0)$; \\
	\lIf{$\textit{Uniform}(0, 1) \le \frac{\exp(-E)}{\exp(-E_{0})}\textcolor{blue}{\frac{\mathbb{P}^{\mathcal{D}}(D^{-1})\exp(\Delta E)}{\mathbb{P}^{\mathcal{D}}(D)}}$}{$p^{H} \leftarrow -p^{H}$}
	\lElse{
		$q^{H}, p^{H}, q^{O} \leftarrow q^{H}_0, p^{H}_{0}, q^{O}_0$
	}
	\KwRet{$q^{H}, p^{H}, q^{O}$}

}
\end{algorithm2e}

\section{Metropolis Augmented Hamiltonian Monte Carlo (MAHMC)}\label{sec:mahmc}

\subsection{Augmenting HMC with MH updates}\label{sec:mahmc-algo}

Motivated by the interpretation of HMC as a deterministic MH proposal (Sec.~\ref{sec:hmc}), for a given distribution $\pi(q^{H}, q^{O}) \propto \exp(-U(q^{H}, q^{O}))$, MAHMC allows more frequent MH updates for $q^{O}$ by combining leapfrog steps for $q^{H}$ and MH updates for $q^{O}$ (including the MH correction) into a single MH proposal, followed by an additional final MH correction.

Formally, for some given step size $\epsilon$ and number of steps $L$, an MAHMC iteration makes use of the leapfrog integrator $\textit{leapfrog}(q, p, \epsilon | U(\cdot, q^{O}))$ (Algo.~\ref{algo:components}) and $N^{O}$ MH proposals $\mathbb{Q}_i(\tilde{q}^{O} | q^{H}, q^{O}), i=1, \ldots, N^{O}$ to construct an MH proposal making $L$ total updates of  $q^{H}, q^{O}$. For a sequence $D \in \left\{ 0, 1, \ldots, N^{O} \right\}^{L}$ of $L$ integers, define $D^{-1} = (D_{L}, D_{L-1}, \ldots, D_{1})$.  Starting from $q^{H}_{0}, q^{O}_{0}$, MAHMC first resamples the momentum $p^{H}_{0}$ from $N(0, I_{n})$, then samples a sequence of $L$ variable updates represented as a sequence $D$ of $L$ integers from some distribution $\mathbb{P}^{\mathcal{D}}(D)$, applies the variable updates one at a time, before making a final MH correction. See Algo.~\ref{algo:mahmc} for a detailed description of an MAHMC iteration.

Critically, the use of MH correction in each MH update serves as a mechanism to prevent MAHMC from deviating too much into low-probability regions. As we empirically verify in Sec.~\ref{sec:exps}, even with the additional MH updates as part of the trajectory, MAHMC can maintain high acceptance probabilities for long-range proposals, similar to HMC. This eliminates the need to balance long HMC trajectories and more frequent other MH updates, and contributes to MAHMC's improved performance over other within-Gibbs alternatives.

\subsection{Connections to M-HMC}\label{sec:mhmc-connection}

\cite{zhou2020mixed, zhou2021erratum} derives M-HMC using auxiliary Hamiltonian dynamics for the discrete variables. We note that the practically useful M-HMC implementation using Laplace momentum is in fact a variant of a special case of MAHMC, where $\mathbb{P}^{\mathcal{D}}$ is implicitly defined using the auxiliary Hamiltonian dynamics and the proposals $\mathbb{Q}_{i}$ are for single discrete variables (given all other discrete and continuous variables). However, M-HMC differs from MAHMC in its use of a persistent kinetic energy $k^{\mathcal{D}}$ for the MH corrections in all MH updates.

\subsection{MAHMC in Algo.~\ref{algo:mahmc} satisfies detailed balance with respect to $\pi(q^{H}, q^{O})$}\label{sec:mahmc-correct}

\begin{proof_sketch} To establish detailed balance, we make use of the concept of \emph{probabilistic paths}, similar to Sec. 2.3 in~\cite{zhou2020mixed, zhou2021erratum}. Starting from $s = (q^{H}_{0}, p^{H}_{0}, q^{O}_{0})$, a \emph{probabilistic path} $\boldsymbol{t}$ contains information about all randomness in an MAHMC iteration (Algo.~\ref{algo:mahmc}):
	\begin{enumerate}[nosep]
		\item The sequence $D$ of $L$ integers specifying which updates to use at each of the $L$ steps.
		\item The actual states $q^{H}_{j}, p^{H}_j, q^{O}_{j}, j=1, \ldots, L$ (before final MH correction) at each step.
		\item The sequence of proposed states $\tilde{q}^{H}_{j}, \tilde{p}^{H}_j, \tilde{q}^{O}_{j}, j=1, \ldots, L$ at each step. 
			\begin{itemize}
				\item If $D_{j} = 0$, $\tilde{q}^{H}_{j}, \tilde{p}^{H}_{j} = \textit{leapfrog}(q^{H}_{j}, p^{H}_{j}, \epsilon | U(\cdot, q^{O}_{j})), \tilde{q}^{O}_{j} = q^{O}_{j-1}$.
				\item If $D_{j} > 0$, $\tilde{q}^{H}_{j}, \tilde{p}^{H}_{j} = q^{H}_{j}, p^{H}_{j}$, $\tilde{q}^{O}_{j}$ is a sample from $\mathbb{Q}_{D_{j}}(\cdot | q^{H}_{j}, q^{O}_{j})$.
			\end{itemize}
		\item The sequence of accept/reject decisions $a_{j} \in \left\{ \texttt{True}, \texttt{False} \right\} , j=1, \ldots, L$ at each step. Note that we always accept ($a_{j} = \texttt{True}$) for leapfrog updates ($D_{j} = 0$).
	\end{enumerate}

We can think of an MAHMC iteration as first sampling a probabilistic path $\boldsymbol{t}$ which brings $s_{0}$ to $s_{L}$, where $s_{j} = (q^{H}_{j}, p^{H}_{j}, q^{O}_{j}), j=1, \ldots, L$, before making an MH correction to either accept $s_{L}$ or reject and return to $s_{0}$. We can \emph{reverse} a probabilistic path $\boldsymbol{t}$ to get probabilistic path  $\boldsymbol{t}^{-1}$ which brings $s_{L}$ back to $s_{0}$. $\boldsymbol{t}^{-1}$ uses the sequence of $L$ updates specified by $D^{-1}$, and reverses the sequence of actual and proposed states (with proper momentum negation) as well as the sequence of accept/rejection decisions. Denote by $\mathbb{P}(\boldsymbol{t} | s_{0})$ the probability of sampling $\boldsymbol{t}$ starting from $s_{0}$, and $\mathbb{P}(s | \boldsymbol{t}), s \in \left\{ s_{0}, s_{L} \right\} $ the MH correction step in MAHMC, we can derive the transition probability of MAHMC as $\mathbb{P}(s' | s) = \sum_{\boldsymbol{t}: s \to  s'} \mathbb{P}(s' | \boldsymbol{t})\mathbb{P}(\boldsymbol{t} | s)$. Define $E(s) = U(q^{H}, q^{O}) + K(p^{H})$. We can establish the desired detailed balance if we can prove $\exp(-E(s))\mathbb{P}(s' | s) = \exp(-E(s'))\mathbb{P}(s | s')$.

We show that $ \mathbb{P}(s_{L} | \boldsymbol{t}) = \min\left\{ 1, \frac{\exp(-E(s_{L}))\mathbb{P}(\boldsymbol{t}^{-1} | s_{L})}{\exp(-E(s_{0}))\mathbb{P}(\boldsymbol{t} | s_{0})} \right\} $ is our desired MH acceptance probability. Note that $\exp(-E(s))\mathbb{P}(s | s) = \exp(-E(s))\mathbb{P}(s | s)$ is trivially true. For $s' \neq s$, 
\begin{align*}
	\exp(-E(s))\mathbb{P}(s' | s) &=  \exp(-E(s))\sum_{\boldsymbol{t}: s \to  s'} \mathbb{P}(s' | \boldsymbol{t})\mathbb{P}(\boldsymbol{t} | s)\\
	&=  \exp(-E(s))\sum_{\boldsymbol{t}: s \to  s'} \min\left\{ 1, \frac{\exp(-E(s'))\mathbb{P}(\boldsymbol{t}^{-1} | s')}{\exp(-E(s))\mathbb{P}(\boldsymbol{t} | s)} \right\}\mathbb{P}(\boldsymbol{t} | s)\\
	&=  \sum_{\boldsymbol{t}: s \to  s'} \min\left\{ \exp(-E(s))\mathbb{P}(\boldsymbol{t} | s), \exp(-E(s'))\mathbb{P}(\boldsymbol{t}^{-1} | s') \right\}\\
	&=  \sum_{\boldsymbol{t}: s' \to  s} \min\left\{ \exp(-E(s))\mathbb{P}(\boldsymbol{t}^{-1} | s), \exp(-E(s'))\mathbb{P}(\boldsymbol{t} | s') \right\}\\
	&= \exp(-E(s'))\mathbb{P}(s | s')
\end{align*}
\vspace*{-15pt}

For $\frac{\mathbb{P}(\boldsymbol{t}^{-1} | s_{L})}{\mathbb{P}(\boldsymbol{t} | s_{0})}$, sampling $D$ contributes $\frac{\mathbb{P}^{\mathcal{D}}(D^{-1})}{\mathbb{P}^{\mathcal{D}}(D)}.$ Leapfrog updates have no randomness and contribute nothing. For $D_{j} > 0$, define $p_{j} = \frac{\exp(-U(\tilde{q}^{H}_j, \tilde{q}^{O}_j))\mathbb{Q}_{D_{j}}(q^{O}_{j} | q^{H}_{j}, \tilde{q}^{O}_{j})}{\exp(-U(q^{H}_j, q^{O}_j))\mathbb{Q}_{D_{j}}(\tilde{q}^{O}_{j} | q^{H}_{j}, q^{O}_{j})} $. If $a_{j} = \texttt{True}$, the MH update contributes $\frac{\mathbb{Q}_{D_{j}}(q^{O}_{j} | q^{H}_{j}, \tilde{q}^{O}_{j})\min\left\{ 1, 1 / p_{j} \right\} }{\mathbb{Q}_{D_{j}}(\tilde{q}^{O}_{j} | q^{H}_{j}, q^{O}_{j})\min\left\{ 1, p_{j} \right\}}  = \frac{\exp(-U(q^{H}_j, q^{O}_j))}{\exp(-U(q^{H}_j, \tilde{q}^{O}_j))}$, which is captured in the $\Delta E$ updates in Algo.~\ref{algo:mahmc}. If $a_{j} = \texttt{False}$, the MH update again contributes nothing (as in both $\boldsymbol{t}$ and  $\boldsymbol{t}^{-1}$ we make the same proposal followed by a rejection). This proves Algo.~\ref{algo:mahmc} uses the correct MH acceptance probability, and establishes the desired detailed balance.
\end{proof_sketch}

\begin{table*}[h!]
\vspace*{-15pt}
\caption{Results on MDC (Sec.~\ref{sec:mdc}) and BLR (Sec.~\ref{sec:blr}). We show ESS per sample per gradients evaluation of $u$ for MDC, and of the potential energy of $\tau, \beta$ for BLR.}
\label{tab:exps}
\centering
\begin{tabular}{cccccc}
\toprule
& MALAwG & HMCwG & MALA-PwG & MALA-PNwG & MAHMCwG \\
\midrule
	MDC  & $1.0\times 10^{-4}$ & $4.62\times 10^{-3}$ & $1.82 \times 10^{-3}$ & $7.38 \times 10^{-3}$ & $\boldsymbol{1.78 \times 10^{-2}}$\\
BLR &  $1.73\times 10^{-3}$ & $7,94\times 10^{-3}$ & $6.66\times 10^{-3}$  & $8.86 \times 10^{-3}$ & $\boldsymbol{9.02 \times 10^{-3}}$\\
\bottomrule
\end{tabular}
\end{table*}

\vspace*{-15pt}

\section{Experiments}\label{sec:exps}

We use the 4 within-Gibbs samplers in Sec.~\ref{sec:within-gibbs} as baselines, and follow the setups of~\cite{neal2020non} when possible. For MALA-based samplers, we alternate between $N^{L}$ leapfrog updates for $q^{H}$ and 1 Gibbs update for $q^{O}$. For HwG, we alternate between HMC iterations and Gibbs updates. Although MAHMC can evolve $q^{H}, q^{O}$ in tandem, to make the comparison most informative, we consider an MAHMC within Gibbs (MAHMCwG) sampler, where we make $N^{U} - 1$ Gibbs updates and $N^{U}N^{L}$ leapfrog updates within each MAHMC iteration, followed by a Gibbs update. The MAHMC iteration schedules the Gibbs and leapfrog updates similarly to MALA-based samplers, alternating between $N^{L}$ leapfrog updates and 1 Gibbs update. This leads to a deterministic $\mathbb{P}^{\mathcal{D}}$ that always proposes the same (symmetric) sequence of updates. Note that the use of a deterministic $\mathbb{P}^{\mathcal{D}}$ makes comparison with~\cite{neal2020non} straightforward, but in general we have more flexibilities in picking $\mathbb{P}^{\mathcal{D}}$. A simple choice is to randomly pick the update to make at each step with a fixed distribution over the different kinds of available updates (i.e. leapfrog and other updates). The fixed distribution allows us to control the relative frequency of the different updates, and the contribution of $\mathbb{P}^{\mathcal{D}}$ to the final acceptance probability can be easily calculated.

We assume gradients evaluation dominates the computation~\citep{neal2020non}, and evaluate performance using effective sample size (ESS) (calculated with~\cite{arviz_2019}) per sample per gradients evaluation. 

Code to reproduce the results is available at \url{https://github.com/StannisZhou/mahmc}.

\subsection{A distribution with mixed discrete and continuous variables (MDC)}\label{sec:mdc}

We consider the distribution in Sec. 5 of~\cite{neal2020non}, where $q^{H}=(u, v), q^{O}=(w_1, \ldots, w_{20})$:
\vspace*{-8pt}
\[
u \sim N(0, 1), v | u \sim N(u, 0.04^{2}), w_{i} | u \sim  \textit{Bernoulli}(1 / (1 + e^{u})), i=1, \ldots, 20
\] 

\paragraph{Correctness} To verify correctness, we compare the histogram of $u$ samples obtained with the 5 samplers, and empirically verify that they match the expected probability density function of $N(0, 1)$ and the samplers are sampling from the right distribution.

\paragraph{Efficiency} We summarize the results in Tab.~\ref{tab:exps}. From the results, we can see that by combining longer HMC trajectories with more frequent Gibbs updates, MAHMCwG is \textbf{3.85x} more efficient than HwG, and \textbf{2.4x} more efficient than MALA-PNwG.

For HwG and MALA-PNwG, we use the optimal hyperparameters reported in Sec. 5 of~\cite{neal2020non} ($L=40$ steps and  step size $\epsilon=0.035$ for HwG, and $N^{L}=10, \epsilon=0.03, \alpha=0.995, \delta = 0.01$ for MALA-PNwG). Our results roughly reproduce the results in~\cite{neal2020non} (MALA-PNwG is 1.6x more efficient than HwG, as opposed to 1.83x reporeted in~\cite{neal2020non}). The small discrepancy can be due to the use of a different metric (ESS of $u$ instead of ESS of $\mathbb{I}(u \in (-0.5, 1.5))$ as in~\cite{neal2020non}, where $\mathbb{I}$ represents the indicator function) and a different way to compute ESS (using~\cite{arviz_2019}).

To make the comparison informative, we use $N^{L}=10$ and $\epsilon=0.03$ for MALAwG and MALA-PwG, and $\alpha=0.995$ for MALA-PwG. We observe that partial momentum refreshment and non-reversible Metropolis accept/reject decisions are indeed beneficial in improving the performance of MALAwG/MALA-PwG. However, only MALA-PNwG outperforms HwG.

The optimal performance of MAHMCwG is achieved with $N^{U}=10$ and $\epsilon=0.04$, i.e. in each MAHMC we make  $N^{U}N^{L}=100$ leapfrog updates and $N^{U - 1}=9$ Gibbs updates (uniformly spreaded). This is far larger than the optimal number of steps $L=40$ for HwG, and demonstrates MAHMC's ability to maintain high acceptance probabilities for long-range proposals that includes MH updates. We additionally test MAHMCwG with $N^{U}=4$ and $\epsilon=0.035$, i.e. making  $N^{U} - 1=3$ additional Gibbs updates on top of HwG, and observe that the normalized ESS increases to  $6.08\times 10^{-3}$. This demonstrates the benefits of more frequent MH updates when we use the same number of leapfrog updates.

\subsection{Bayesian Logistic Regression (BLR) with conjugate prior}\label{sec:blr}
We apply Bayesian Logistic Regression (BLR) to the breast cancer wisconsin dataset\footnote{Available on the \href{https://archive.ics.uci.edu/ml/datasets/Breast+Cancer+Wisconsin+(Diagnostic)}{UCI Machine Learning Repository}. Accessed through \href{https://scikit-learn.org/stable/modules/generated/sklearn.datasets.load_breast_cancer.html}{sklearn}.}, consisting of $569$ pairs of $30$ dimensional features and targets $y_i \in \left\{ 0, 1 \right\} $. We standarize the features and append $1$ at the end to get  $x_{i}\in \mathbb{R}^{31}$, and specify BLR with conjugate prior as
\vspace*{-2pt}
\[
\tau \sim \textit{Gamma}(1.0, scale=100), \beta \sim N(0, \frac{1}{\tau} I_{31}), y_{i}\sim \textit{Bernoulli}(\textit{sigmoid}(x_{i}^{T}\beta)), i=1, \ldots, 569
\vspace*{-3pt}
\] 
We consider sampling from the posterior $\mathbb{P}(\tau, \beta | X, y)$, where $q^{H}=\beta$ and $q^{O}=\tau$, and we make Gibbs updates for $\tau$ using $\mathbb{P}(\tau | \beta, X, y)$. 

\paragraph{Correctness} To verify the correctness of the samplers, we first apply the samplers to only the prior distribution
\[
\tau \sim \textit{Gamma}(1.0, scale=100), \beta \sim N(0, \frac{1}{\tau} I_{31})
\] 
and similarly verify the marginal distribution of $\tau$ is indeed  $\textit{Gamma}(1.0, scale=100)$ by looking at the histogram of $\tau$ samples. We additionally verify that all samplers give rough the same posterior means for  $\beta$, and we can classify the 569 training data points to an accuracy of $98.77\%$ using posterior samples from the 5 different samplers.

\paragraph{Efficiency} We fix $N^{L} = 5$, and conduct a grid search in
\[
N^{U}\in \left\{ 2, 3, 4, 5 \right\}, \epsilon \in \left\{ 0.07, 0.08, 0.09, 0.10, 0.11 \right\}, \alpha \in \left\{ 0.9, 0.95, 0.98, 0.99 \right\}
\]
and $\delta \in \left\{ 0.005, 0.01, 0.015, 0.03 \right\} $, and $L \in \left\{ 10, 15, 20, 25 \right\} $ for HwG. 

Optimal performance is achieved with $\epsilon=0.11$ for MALAwG,  $L=10, \epsilon=0.09$ for HwG, $\epsilon=0.09, \alpha=0.9$ for MALA-PwG, $\epsilon=0.1, \alpha=0.9, \delta=0.015$ for MALA-PNwG, and $N^{U}=2, \epsilon=0.1$ for MAHMCwG.

For this example, MAHMCwG only slightly outperforms MALA-PNwG. However, we comment that MAHMCwG is easier to tune, as we essentially only need to tune the step size and number of steps, same as in HwG, and using the same setups from HwG usually already gives good performance. However, for MALA-PwG, we need to tune $\alpha$ and $\delta$, which can have significant impacts on performance. For example, for the distribution in~\ref{sec:mdc}, grid search is done for $\alpha\in \left\{ 0.98, 0.99, 0.995, 0.9975, 0.9985, 0.999 \right\} $ in~\cite{neal2020non}. However, in our experiments we empirically observe that all these values give poor performance, and we have to reduce $\alpha$ to  $0.9$ to get performance comparable with MAHMCwG, suggesting the potential challenges in tuning $\alpha$.

\subsection{Additional verification of correctness}

Since we only used Gibbs updates (which always accept) in our experiments, we additionally verify the correctness of MAHMC by modifying the \href{https://github.com/StannisZhou/mixed_hmc/blob/master/scripts/simple_gmm/test_naive_mixed_hmc.py}{script} used in~\cite{zhou2020mixed, zhou2021erratum} to make random walk MH updates within HMC for the 1D GMM example using MAHMC. See \url{https://github.com/StannisZhou/mahmc/blob/main/simplified_mixed_hmc.py} for the updated script.

\bibliography{refs}

\end{document}